\documentclass[
 amsmath,amssymb,
 reprint,%
]{revtex4-1}

\usepackage{graphicx}
\usepackage{dcolumn}
\usepackage{bm}
\usepackage{indentfirst}
\usepackage[utf8]{inputenc}
\usepackage[T1]{fontenc}
\usepackage{mathptmx}
\usepackage{etoolbox}
\usepackage{subfigure}
\usepackage{amsmath}

\makeatletter
\def\@email#1#2{%
 \endgroup
 \patchcmd{\titleblock@produce}
  {\frontmatter@RRAPformat}
  {\frontmatter@RRAPformat{\produce@RRAP{*#1\href{mailto:#2}{#2}}}\frontmatter@RRAPformat}
  {}{}
}%
\renewcommand{\section}{\@startsection{section}{1}{0mm}
  {-\baselineskip}{0.2\baselineskip}{\bf\leftline}}
\renewcommand{\subsection}{\@startsection{subsection}{1}{0mm}
  {-\baselineskip}{0.2\baselineskip}{\bf\leftline}}
\makeatother

\begin{document}

\preprint{AIP/123-QED}

\title[Sample title]{Novel Sample Stages for Detecting Magnetic Feedback Due to \\Superconducting Transition Based on the Mirror Image Method }
\author{Ziyan Li}
\author{Mengbo Guo}
\affiliation{ 
Physics Department, Wuhan University, Wuhan 430072, China
}%



\begin{abstract}
In traditional detection of the magnetic signal\cite{yusa2021probabilistic,tuzlukov2001signal,pastore1974signal} in the experiment, complicated electronic devices are usually set up to explore the feedback from the superconducting transition due to the external magnetic field. This kind of direct detection will be more time-consuming, especially in experiments with strict requirements like low temperature and high vacuum. To make the signal detection more convenient to realize, we design a novel sample stage with functions of both placement and acceptance of the magnetic field based on the mirror image method\cite{xifre2013mirror,yakushev2013theory}. This device consists of two parts, drive coils and receiver coils. The drive coils apply the initial magnetic field to the superconducting material while the receiver coils are used to collect the magnetic signal feedback. With only the power supply and a signal amplifier needed to be connected to the coils, our device makes it more convenient to finish magnetic signal detection. 
Another highlight of our design is using first-principles numerical simulation to help accurately determine the size and parameters of the device. With these calculated parameters, we build a 3D model of the device using the software SOLIDWORKS.
\end{abstract}

\maketitle
\section{\label{sec:level1}Introduction}

\begin{figure*}[htbp]
   \centering
   \includegraphics[scale=0.22]{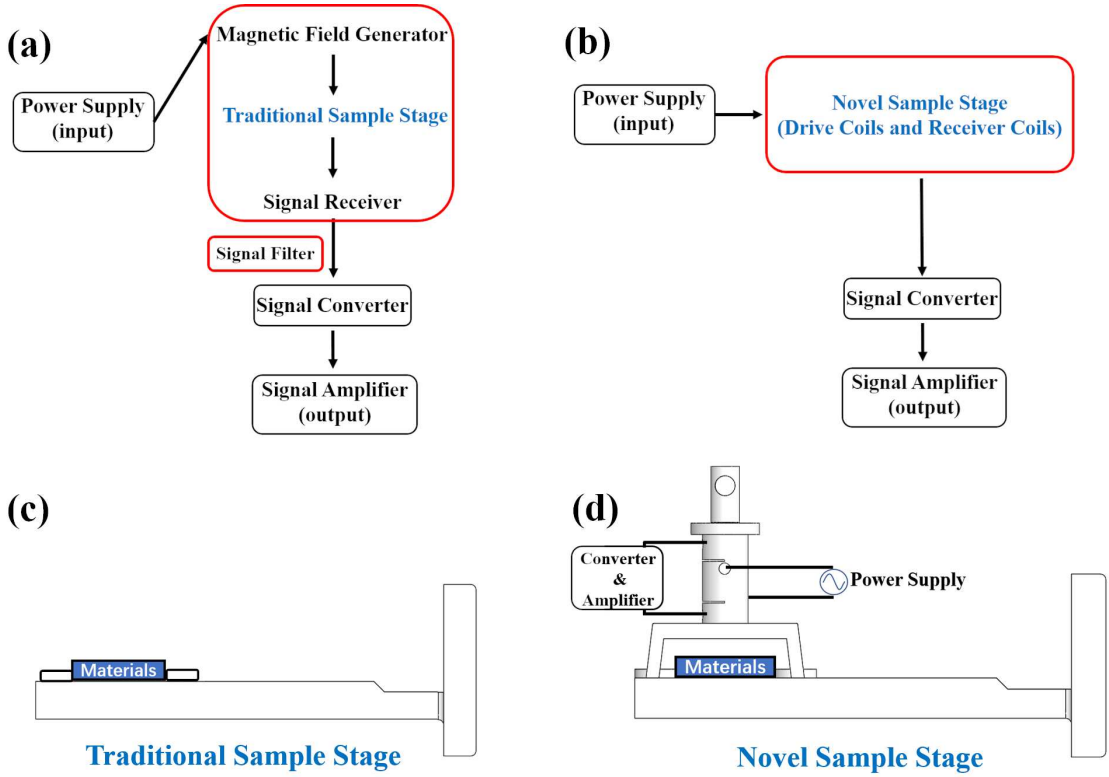}
    \caption{(a) The traditional detection process of the magnetic signal. (b) The novel detection process of magnetic signal with the novel sample stage. (c) The traditional sample stage for holding sample materials. (d) The novel sample stage with drive coils and receiver coils connected with the outside power supply and the signal amplifier. }
    \label{fig:2}
\end{figure*}

In experiments to explore the magnetic properties of superconducting materials, ultra-high vacuum and low temperature are often necessary conditions. The traditional detection\cite{zhao2021brief,tian2022pfmd} of magnetic signals under such conditions will always require complicated electronic devices including an initial signal generator, a traditional sample stage, a receiver and a signal filter, as shown in the Fig.~\ref{fig:2}(a). Both applying and receiving the magnetic signal will be usually time-consuming with such a series of devices. To make the detection more convenient, we modified the sample stage to be equipped with drive coils and receiver coils. Therefore, we could simplify the detection process with this novel sample stage as Fig.~\ref{fig:2}(b) shows. Only two wires are needed to connect our designed device with the outside power supply and the signal amplifier \cite{brkovic1994novel}.

The sample material to be tested is placed on the novel sample stage in Fig.\ref{fig:2}(d) and sent into the chamber. Once the sample stage is in place, the interior of the chamber must be isolated from the outside world. Therefore, a well-designed stage with an accurate size is important to ensure that the feedback magnetic field is large enough to be measured. To find the appropriate parameters of our device model, we refer to the mirror image method\cite{Emig_2008,xifre2013mirror}. In addition, we apply the Messner Effect\cite{essen2012meissner} theory to numerically simulate the magnetic signal feedback due to the superconducting transition. The highlight of our simulation is that 
we could separate initial and feedback magnetic signals so it will be easier for conducting the signal-filtering process in the experiment.

We show the 3D model diagram of the device using the software SOLIDWORKS. Moreover, to reduce the error caused by the boundary effect of magnetic field\cite{iverson1995mapping,naitou1979boundary}, we analyze in FIG.~\ref{fig:6} to discuss the critical size of the sample stage, which is large enough that the boundary effect can be ignored.

\section{\label{sec:level2}NUMERICAL SIMULATION MODEL}

\par In order to simultaneously apply the magnetic field and detect the magnetic field feedback, we place the drive coils and receiver coils above the sample materials. The simulation of the initial magnetic field applied by the drive coils can be realized with the Biot-Savart law\cite{Charitat_2003,oliveira2001biot}. The feedback magnetic signal due to superconducting transition is numerically simulated referring to the mirror image method.

\subsection{\label{sec:2_A}Drive Coils to Apply the Magnetic Field}

\begin{figure}[htbp]
   \centering
   \includegraphics[scale=0.18]{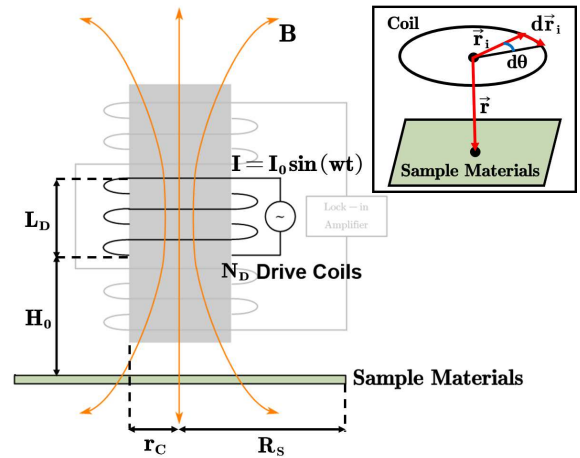}
   \caption{The drive coils are connected to the power supply and will apply the initial magnetic field to the sample material. The inset shows the definition of some important parameters in equation~(\ref{equ:1}).}
   \label{fig:1}
\end{figure}

In FIG.\ref{fig:1}, $\rm L_D$ is the length of drive coils while $\rm H_0$ stands for the distance between the bottom of drive coils and sample materials. $\rm r_{C}$ and $\rm R_S$ separately represent the radius of drive coils and that of sample materials. The drive coils are connected to the external AC power supply. The current in the coils satisfies $\rm I=I_0 sin(\omega t)$, where $\rm \omega = 2\pi f$ is the angular frequency. By applying the Biot-Savart law, we obtain the magnitude of the magnetic field $\rm B(h)$ generated by each coil at a point on its axis at a distance $\rm h$.

\begin{equation}
  \rm \boldsymbol{B}(h) =\frac{\mu_0}{4\pi}\oint\frac{Id\overrightarrow{\boldsymbol{r_i}}\times(\overrightarrow{\boldsymbol{r}}-\overrightarrow{\boldsymbol{r_i}})}{|\overrightarrow{\boldsymbol{r}}-\overrightarrow{\boldsymbol{r_i}}|^3} \\
 =\frac{\mu_0 I}{4\pi}\sum_{i=0}^{N_{\theta}}\frac{r_{C}d\theta \overrightarrow{\boldsymbol{e_{i}}}\times(-\overrightarrow{\boldsymbol{r_i}})}{(h^2+r_{C}^2)^{\frac{3}{2}}}
 \label{equ:1}   
\end{equation}

 The equation~(\ref{equ:1}) represents for the discretization of Biot-Savart law\cite{1128246,arnason1999consistent,BOSSAVIT2005105}, and the circumference angle $\rm \theta$ is discretized into $\rm N_{\theta}$ parts. We sum the magnetic field strengths $\rm B_{D}$ generated by $\rm N_D$ drive coils on the axis to obtain the magnetic field distribution on the axis of the drive coils.

\begin{equation}
 \rm \boldsymbol{B}_{D}(h_0,N_{D}) = \sum_{i=0}^{N_{D}-1}\boldsymbol{B}(h_0+i\Delta L)
 \label{equ:2}
\end{equation}

$\rm h_0$ refers to the distance between the bottom of the drive coil and the point on the axis. For example, if we consider the distance from the center of the sample then $\rm h_0 = H_0$ (See FIG.~\ref{fig:1}). The distance between adjacent coils $\rm \Delta L$ in equation(\ref{equ:2}) is determined by $\rm L_{D}/N_{D}$.

\subsection{\label{sec:2_B} Receiver Coils and the Mirror Image Method}

The magnetic field generated by the drive coils will pass the sample material. With a temperature lower than the superconducting critical temperature, a surface current will form on the sample according to the Meissner effect\cite{bardeen1955theory,schafroth1958remarks}, and the magnetic field generated by the surface current will change the magnetic field in the space. We use mirror coils to simulate the magnetic field feedback due to the surface current. 

\begin{figure}[htbp]
   \centering
   \includegraphics[scale=0.18]{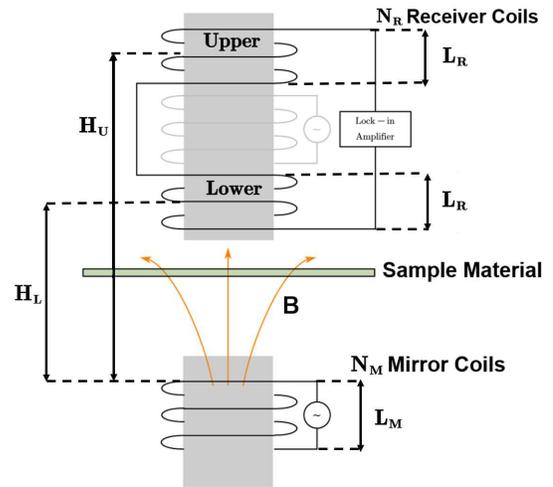}
   \caption{\label{fig:3} The receiver coils are connected to the amplifier to amplify the feedback magnetic signal due to the superconducting transition. The mirror coils are the mirror image of the drive coils relative to the sample stage.}
   \label{fig:3}
\end{figure}

 In FIG.~\ref{fig:3}, $\rm H_U$ stands for the distance between the middle of upper receiver coils and the top of mirror coils, while $\rm H_L$ is the distance between the middle of lower receiver coils and mirror coils top. $\rm L_R$ and $\rm L_M$ separately represent the length of lower/upper receiver coils and mirror coils
 
We divide the receiver coils into two parts, the upper and the lower coils. Both these two parts have the same length $\rm L_R$ and the same number of coils $\rm N_R$. The drive coils are set in the middle of these two parts so as to eliminate the influence of the magnetic field generated in the receiver coils by the drive coils because the produced electromotive force in the upper and lower receiver coils will be equal and reversed. In order to simplify the calculation, We assume that the magnetic field generated by the mirror coils inside the receiver coils is constant and can be replaced by the magnetic field generated by the mirror coils in the middle of the receiver coils. 

Based on the Faraday electromagnetic induction law\cite{Giuliani_2008,galili2006teaching}, we obtain the electromotive force difference $\rm \Delta U$ generated by the mirror coils at the receiver coils in the equation~(\ref{equ:3}).

\begin{widetext}
\begin{equation}
  \begin{aligned}
   && \rm \Delta U & \rm =|U_{U}-U_{L}| &\\
   && & \rm =|-N_R\pi r_C^2\frac{dB_D(H_L,N_D)-dB_D(H_U,N_D)}{dt}| &\\
   && & \rm = N_{R}f \pi r_{C}^2 \frac{\mu_0 I_{rms}}{2} |\sum_{i=0}^{N_{D}-1} \Big \{ \sum_{j=0}^{N_{\theta}}\frac{r_{C}d\theta [\overrightarrow{\boldsymbol{e_j}}\times(-\overrightarrow{\boldsymbol{r_j}})]_z}{[(H_{U}+\frac{i L_{D}}{N_{D}})^2+r_{C}^2]^{\frac{3}{2}}}-\sum_{j=0}^{N_{\theta}}\frac{r_{C}d\theta [\overrightarrow{\boldsymbol{e_j}}\times(-\overrightarrow{\boldsymbol{r_j}})]_z}{[(H_{L}+\frac{iL_{D}}{N_{D}})^2+r_{C}^2]^{\frac{3}{2}}}\Big \}|
  \end{aligned}
\label{equ:3}
\end{equation}
\end{widetext}

 The potential difference shown in equation~(\ref{equ:3}) can be amplified by the amplifier\cite{1328930,guide18active} for subsequent analysis. The existing instrument can measure the potential difference greater than 100nV. That is to say, as long as the measured potential difference is greater than 100nV, it can satisfy the needs of the experiment.

\subsection{\label{sec:2C}Determination of Parameters in the Model}
In equation~(\ref{equ:3}), there are many parameters set to be constant, such as $\rm N_{R}$, $\rm N_{D}$, $\rm r_{C}$ and $\rm \Delta L$ in this paper. This requires us to fix these parameters when designing the model. To ensure that the voltage difference received by the receiver coils  $\rm \Delta U$  during the experiment is large enough, we first decide some parameters according to the size of other experimental devices. For example, the length $\rm L_D$ and radius $\rm r_C$ of the drive coils should match the sample stage's size. Therefore, we select some appropriate values for parameters in
TABLE~\ref{tab:1}.

\begin{center}
\begin{table}[htbp]
    \centering
    \caption{\label{tab:1} Other parameters except $\rm N_{R}$ and $\rm N_{D}$, which are referred to the specifications and sizes of the specific experimental equipment. $\rm I_{rms}$ is the effective value of the alternating current in the drive coils.}
    \setlength{\tabcolsep}{0.8mm}{
    \begin{tabular}{|c c c c c c c |} 
         \hline
         $\rm f \ (kHz)$ & $ \rm r_{C}$ (mm) & $\rm I_{rms}\ (\mu A)$ & $\rm H_{L}$ (mm) & $\rm H_{U}$ (mm) & $\rm L_{D} \ (mm)$ & $\rm N_{\theta}$  \\  [0.2ex]
         \hline\hline
         100  & 2 & 500 & 17.5 & 30.5 & 8 & 100\\ [0.2ex] 
         \hline
     \end{tabular}}
     \label{tab:1}
\end{table}
\end{center}

\begin{figure}[h]
   \centering
   \includegraphics[scale=0.15]{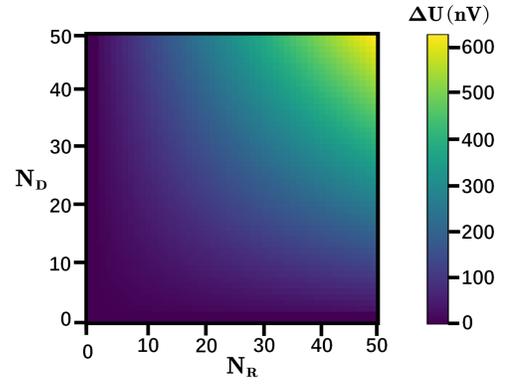}
   \caption{The voltage difference $\rm \Delta U$ is a function of $\rm N_D$ and $\rm N_R$. $\rm \Delta U$ will exceed 600nV when $\rm N_{R}>50$, $\rm N_{D}>50$.}
    \label{fig:4}
\end{figure}

However, some parameters like the number of coils are flexible and will also affect the detection results. To find the appropriate $\rm N_{D}$ and $\rm N_{R}$ to maximize $\rm \Delta U$, we use the equation~(\ref{equ:3}) to calculate the potential difference as a function of $\rm N_{D}$ and $\rm N_{R}$. 

FIG~\ref{fig:4} shows us the results of the numerical calculation. It can be seen that the calculated voltage difference increases with the increasing $\rm N_{D}$ and $\rm N_{R}$. However, the number of coil turns should not be too large, if not so the distance between the coils will be would be smaller than the size of the coil itself, which is impossible. We can take $\rm N_{D}$=40, $\rm N_{R}$=30 so the corresponding $\rm \Delta U$ is 314.06nV. It is already greater than the detectable 100nV, so this turns ratio is acceptable.
\subsection{3D Modeling of the Device}

We can use SOLIDWORKS to draw the 3D model\cite{schilling2019parametric,akin2010finite} diagram to have an intuitive understanding of the model. The scale of the drawing is not drawn according to the real scale, which is for the convenience of display.

\begin{figure*}[htbp]
   \centering
   \includegraphics[scale=0.24]{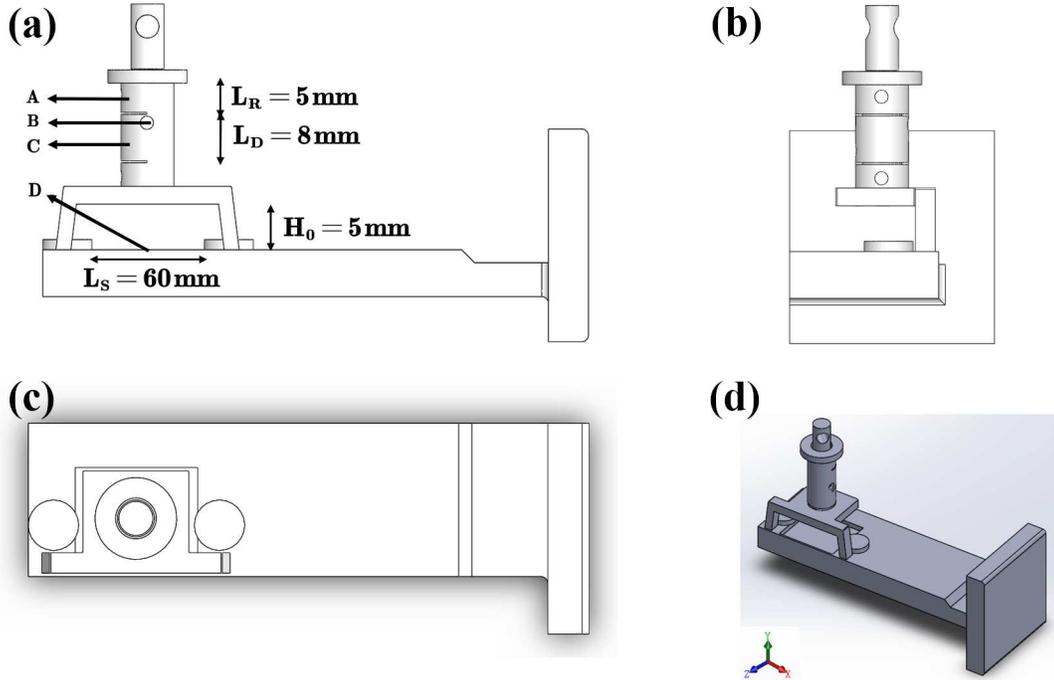}
    \caption{The 3D model is not showed in the real scale for the convenience of display. (a) The front view of the model. A is the compartment where the receiver coils are placed, B is the hole used for connecting the external power supply, C is the compartment where the drive coils are placed and D is the sample stage holding the sample materials. (b) The left view. (c) The top view. (d) The view of the whole model.}
       \label{fig:5}
\end{figure*}

From the main view in FIG~\ref{fig:5}(a), we can see that the entire model is mainly composed of two parts: a movable cylindrical barrel for loading coils and a sample stage for placing samples. The cylindrical barrel consists of three layers, the middle layer is used to place the drive coils, the upper and lower layers are used to place the receiver coils, and each layer has holes to connect with the external driving circuit or receiving circuit. Just below the cylindrical barrel is the place to hold the sample, and the distance $\rm H_0$ from the sample to the bottom of the cylindrical barrel can be changed as required.

\section{The Error Analysis}

\begin{figure}[htbp]
   \centering
   \includegraphics[scale=0.13]{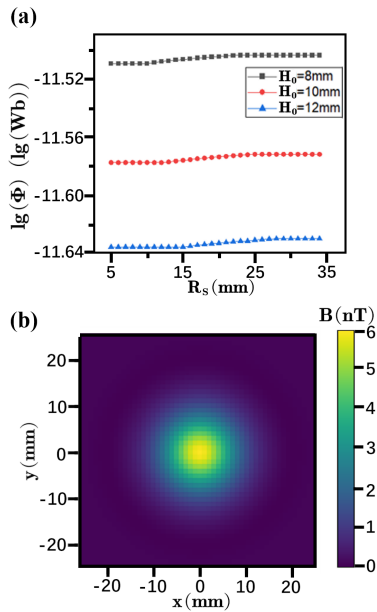}
    \caption{(a) Convergence of magnetic flux $\rm \Phi$ in the sample material with $\rm H_0$=8mm, 10mm and 12mm. (b) The initial magnetic field B is applied in the sample plane with height $\rm H_0$=10mm.}
    \label{fig:6}
\end{figure}

Usually, further experimental validation could be a good way to prove the reliability of our model design. However, we would like to suggest it as a topic for further research due to a number of complications in experimental detection. Other alternative ways are applied to improve the reliability in this paper. We maximize the fault tolerance\cite{avizienis1984fault} and analyze the boundary effect to reduce possible errors in our model design. This kind of error analysis\cite{bertoni2003error} will also help improve our design so that it can be well-applied to the magnetic signal detection experiments.

\subsection{The Fault Tolerance Maximization}

When a device is applied to the experimental detection, many parameters will not be able to be revised. Therefore, we should always improve our design to maximize fault tolerance. For example, Fig.~\ref{fig:6}(a) shows that the magnetic flux  $\rm \Phi$ will converge with different values of $\rm R_S$ for different $\rm H_0$. To maximize the fault tolerance, we should choose the maximum $\rm R_S$=30mm to ensure $\rm \Phi$ will converge for different $\rm H_0$.

\subsection{The Boundary effect}

An important prerequisite for the mirror image method is that we can ignore the boundary effect of the sample\cite{8653466,5352342,EVANS2005691,yu1991modeling}. Therefore, we need to estimate the critical length of the stage to ensure that the magnetic field outside the sample stage will be negligible. In FIG~\ref{fig:6}(b), the magnetic field will decrease to 0 when the position is far from the center of the sample stage, which means that the magnetic field will be negligible compared with the central value when the sample size is larger than 30mm. In fact, there are other numerical simulation methods(details in Appendix~\ref{Appendix:B}), but they all need to consider the boundary effects of electromagnetic fields.

\section{Conclusion}
To make the detection of the magnetic signal due to the superconducting transition more convenient, we design a novel sample stage based on the mirror image method. This design adds the drive coils and receiver coils to the traditional sample stage, which reduces many electronic devices and simplifies the detection process. The numerical simulation skills are applied to find the most suitable size parameters for the sample stage. Moreover, we build a 3D model diagram to help understand the 3D structure of our device. We further explore the possible reasons for design errors and try to find a stage size which is large enough to eliminate the boundary effect.

\hspace*{\fill} \\ 
\noindent \textbf{\normalsize ACKNOWLEDGMENTS}
\par We wish to acknowledge the support of Professor Jing Shi from Wuhan University, China, who encourages us to finish and perfect this project. We also wish to acknowledge the guidance of Professor Yuanbo Zhang, Master Zhiwei Huang and Master Hengsheng Luo from Fudan University, China, providing the experimental environment and gave some important suggestions.
\hspace*{\fill} \\ 

\noindent \textbf{\normalsize AUTHOR DECLARATIONS}

\noindent \textbf{Conflict of Interest}
\par The authors declare no conflicts of interest.

\noindent \textbf{Author Contributions}

\par Ziyan Li is the major contributor and Mengbo Guo is the minor contributor.

\hspace*{\fill} \\ 
\noindent \textbf{\normalsize DATA AVAILABILITY STATEMENT}
\par The data that support the findings of
this study are available within the
article [and its supplementary material].

\appendix

\section{\label{Appendix:A}Meissner Effect and London Equation}
\begin{figure}[h]
   \centering
   \includegraphics[scale=0.22]{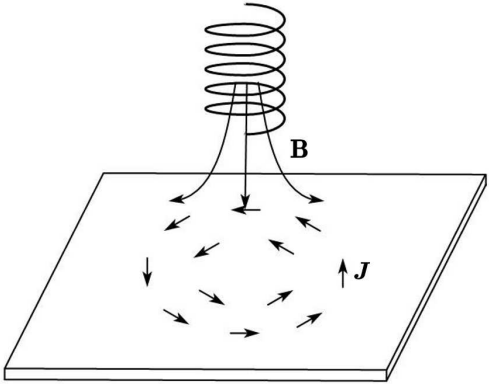}
   \caption{The surface current generated by the magnetic field applied to the superconducting sample materials.}
   \label{fig:7}
\end{figure}

The Meissner effect refers to the phenomenon\cite{anderson1964superconductivity} that the magnetic field will maintain zero inside the superconducting materials. The London equation\cite{kaloper2017london} is a relatively classic and simple explanation model to explain this phenomenon.

This model assumes that electrons in the superconducting state can be divided into two types: electrons in the normal state and electrons in the superconducting state. The surface current density of the material can be related to the corresponding magnetic vector potential: 

\begin{equation}
\label{equ:A1}
\rm \boldsymbol{J}=-\frac{n_s e^2}{m_e}\boldsymbol{A}
\end{equation}

Take the curl for quantities on both sides and refer to Maxwell's equation\cite{lax1976maxwell,ASSOUS1993222}. Then we can get the relation between current density and magnetic field:
\begin{equation}
\label{equ:A2}
\rm \nabla \times \boldsymbol{J} = - \frac{n_s e^2}{m_e}\boldsymbol{B}
\end{equation}

\begin{equation}
\label{equ:A3}
\rm \nabla \times \boldsymbol{B} = \mu_0 \boldsymbol{J}
\end{equation}

Combining all the above formulas to obtain the magnetic field:

\begin{equation}
\label{equ:A4}
\rm \boldsymbol{B}(x) = \boldsymbol{B_0} exp(-\frac{x}{\lambda})
\end{equation}

Equation (\ref{equ:A4}) shows that the magnetic field decreases exponentially, indicating that the magnetic field cannot penetrate the surface of the sample. As shown in FIG.~\ref{fig:7}, the surface current\cite{gurevich1993photomagnetism,gurevich1992photomagnetism} can be used to calculate the magnetic field feedback generated by the sample. This approach is essentially consistent with the mirror method.

\section{\label{Appendix:B}Numerical Simulation of Annular Current}

\begin{figure}[htbp]
   \centering
   \includegraphics[scale=0.14]{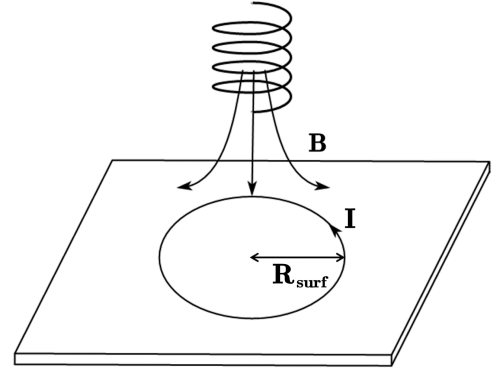}
   \caption{ The ring current is used to be in place of the surface current. $\rm R_{surf}$ is the radius of the ring current.}
   \label{fig:8}
\end{figure}
This is also a method similar to using surface currents to simulate feedback magnetic fields, but unlike Appendix \ref{Appendix:A}, this method replaces the surface currents with a loop of toroidal currents. The feedback of the superconducting material can be calculated if we know about the value of the current. The method of confirming the magnitude of the current is making the sum of the magnetic fluxes generated by the annular current inside it equal to the magnetic fluxes generated by the external magnetic field.

Similar to equation~(\ref{equ:1}), taking the center of the ring current as the origin, we can find the vertical component of the magnetic field generated by the ring current at its interior $\overrightarrow{\boldsymbol{r}}$:

\begin{equation}
\rm \boldsymbol{B}_{surf}(\overrightarrow{\boldsymbol{r}})
 =\frac{\mu_0 I}{4\pi}\sum_{i=0}^{N_{\theta}}\frac{|\overrightarrow{\boldsymbol{r}}|d\theta \overrightarrow{\boldsymbol{e_{\theta_i}}}\times(\overrightarrow{\boldsymbol{r}}-\overrightarrow{\boldsymbol{r_i}})}{|\overrightarrow{\boldsymbol{r}}-\overrightarrow{\boldsymbol{r_i}}|^3}
 \label{equ:B1}
\end{equation}

Integrating the magnetic field inside the ring current can get the magnetic flux $\rm \Phi_{surf}$ generated by the ring current:

\begin{equation}
 \rm \Phi_{surf} = \int_{0}^{R_{surf}} |B_{surf}|_z 2\pi rdr
 \label{equ:B2}
\end{equation}

\begin{figure}[htbp]
   \centering
   \includegraphics[scale=0.14]{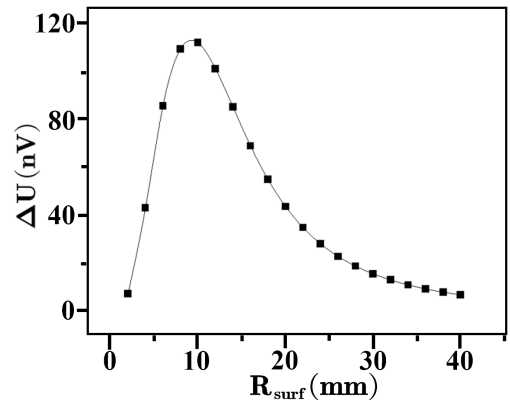}
   \caption{The potential difference $\rm \Delta U$ is generated by the ring current. It's a function of the ring radius $\rm R_{surf}$.}
   \label{fig:9}
\end{figure}

As for the sum of the magnetic fluxes $\rm \Phi_{Dri}$ generated by the driving coil inside the annular current, it is not difficult to obtain it with equation (\ref{equ:2}). Therefore, we can obtain the equation, $\rm \Phi_{surf} + \Phi_{Dri} = 0$ to calculate the magnitude of the annular current. When we take the same parameters as in Table~\ref{tab:1} and choose different radii of the ring current, the potential difference calculated by the method of annular current is shown in FIG.~\ref{fig:9}. It can be seen from FIG.~\ref{fig:9} that $\rm \Delta U$ calculated by the ring current will vary with different radius $\rm R_{surf}$ taken, indicating that this is not a suitable method. What's more,
the maximum value is about 115nV, which is lower than the 314.06nV measured by the mirror method.

\bibliography{reference}

\providecommand{\noopsort}[1]{}\providecommand{\singleletter}[1]{#1}%
\begin{thebibliography}{37}%
\makeatletter
\providecommand \@ifxundefined [1]{%
 \@ifx{#1\undefined}
}%
\providecommand \@ifnum [1]{%
 \ifnum #1\expandafter \@firstoftwo
 \else \expandafter \@secondoftwo
 \fi
}%
\providecommand \@ifx [1]{%
 \ifx #1\expandafter \@firstoftwo
 \else \expandafter \@secondoftwo
 \fi
}%
\providecommand \natexlab [1]{#1}%
\providecommand \enquote  [1]{``#1''}%
\providecommand \bibnamefont  [1]{#1}%
\providecommand \bibfnamefont [1]{#1}%
\providecommand \citenamefont [1]{#1}%
\providecommand \href@noop [0]{\@secondoftwo}%
\providecommand \href [0]{\begingroup \@sanitize@url \@href}%
\providecommand \@href[1]{\@@startlink{#1}\@@href}%
\providecommand \@@href[1]{\endgroup#1\@@endlink}%
\providecommand \@sanitize@url [0]{\catcode `\\12\catcode `\$12\catcode
  `\&12\catcode `\#12\catcode `\^12\catcode `\_12\catcode `\%12\relax}%
\providecommand \@@startlink[1]{}%
\providecommand \@@endlink[0]{}%
\providecommand \url  [0]{\begingroup\@sanitize@url \@url }%
\providecommand \@url [1]{\endgroup\@href {#1}{\urlprefix }}%
\providecommand \urlprefix  [0]{URL }%
\providecommand \Eprint [0]{\href }%
\providecommand \doibase [0]{http://dx.doi.org/}%
\providecommand \selectlanguage [0]{\@gobble}%
\providecommand \bibinfo  [0]{\@secondoftwo}%
\providecommand \bibfield  [0]{\@secondoftwo}%
\providecommand \translation [1]{[#1]}%
\providecommand \BibitemOpen [0]{}%
\providecommand \bibitemStop [0]{}%
\providecommand \bibitemNoStop [0]{.\EOS\space}%
\providecommand \EOS [0]{\spacefactor3000\relax}%
\providecommand \BibitemShut  [1]{\csname bibitem#1\endcsname}%
\let\auto@bib@innerbib\@empty
\bibitem [{\citenamefont {Yusa}\ \emph {et~al.}(2021)\citenamefont {Yusa},
  \citenamefont {Song}, \citenamefont {Iwata}, \citenamefont {Uchimoto},
  \citenamefont {Takagi},\ and\ \citenamefont {Moroi}}]{yusa2021probabilistic}%
  \BibitemOpen
  \bibfield  {author} {\bibinfo {author} {\bibfnamefont {N.}~\bibnamefont
  {Yusa}}, \bibinfo {author} {\bibfnamefont {H.}~\bibnamefont {Song}}, \bibinfo
  {author} {\bibfnamefont {D.}~\bibnamefont {Iwata}}, \bibinfo {author}
  {\bibfnamefont {T.}~\bibnamefont {Uchimoto}}, \bibinfo {author}
  {\bibfnamefont {T.}~\bibnamefont {Takagi}}, \ and\ \bibinfo {author}
  {\bibfnamefont {M.}~\bibnamefont {Moroi}},\ }\href@noop {} {\bibfield
  {journal} {\bibinfo  {journal} {Nondestructive Testing and Evaluation}\
  }\textbf {\bibinfo {volume} {36}},\ \bibinfo {pages} {1} (\bibinfo {year}
  {2021})}\BibitemShut {NoStop}%
\bibitem [{\citenamefont {Tuzlukov}(2001)}]{tuzlukov2001signal}%
  \BibitemOpen
  \bibfield  {author} {\bibinfo {author} {\bibfnamefont {V.~P.}\ \bibnamefont
  {Tuzlukov}},\ }\href@noop {} {\emph {\bibinfo {title} {Signal detection
  theory}}}\ (\bibinfo  {publisher} {Springer Science \& Business Media},\
  \bibinfo {year} {2001})\BibitemShut {NoStop}%
\bibitem [{\citenamefont {Pastore}\ and\ \citenamefont
  {Scheirer}(1974)}]{pastore1974signal}%
  \BibitemOpen
  \bibfield  {author} {\bibinfo {author} {\bibfnamefont {R.}~\bibnamefont
  {Pastore}}\ and\ \bibinfo {author} {\bibfnamefont {C.}~\bibnamefont
  {Scheirer}},\ }\href@noop {} {\bibfield  {journal} {\bibinfo  {journal}
  {Psychological Bulletin}\ }\textbf {\bibinfo {volume} {81}},\ \bibinfo
  {pages} {945} (\bibinfo {year} {1974})}\BibitemShut {NoStop}%
\bibitem [{\citenamefont {Xifre-Perez}\ \emph {et~al.}(2013)\citenamefont
  {Xifre-Perez}, \citenamefont {Shi}, \citenamefont {Tuzer}, \citenamefont
  {Fenollosa}, \citenamefont {Ramiro-Manzano}, \citenamefont {Quidant},\ and\
  \citenamefont {Meseguer}}]{xifre2013mirror}%
  \BibitemOpen
  \bibfield  {author} {\bibinfo {author} {\bibfnamefont {E.}~\bibnamefont
  {Xifre-Perez}}, \bibinfo {author} {\bibfnamefont {L.}~\bibnamefont {Shi}},
  \bibinfo {author} {\bibfnamefont {U.}~\bibnamefont {Tuzer}}, \bibinfo
  {author} {\bibfnamefont {R.}~\bibnamefont {Fenollosa}}, \bibinfo {author}
  {\bibfnamefont {F.}~\bibnamefont {Ramiro-Manzano}}, \bibinfo {author}
  {\bibfnamefont {R.}~\bibnamefont {Quidant}}, \ and\ \bibinfo {author}
  {\bibfnamefont {F.}~\bibnamefont {Meseguer}},\ }\href@noop {} {\bibfield
  {journal} {\bibinfo  {journal} {ACS nano}\ }\textbf {\bibinfo {volume} {7}},\
  \bibinfo {pages} {664} (\bibinfo {year} {2013})}\BibitemShut {NoStop}%
\bibitem [{\citenamefont {Yakushev}(2013)}]{yakushev2013theory}%
  \BibitemOpen
  \bibfield  {author} {\bibinfo {author} {\bibfnamefont {E.}~\bibnamefont
  {Yakushev}},\ }in\ \href@noop {} {\emph {\bibinfo {booktitle} {Advances in
  Imaging and Electron Physics}}},\ Vol.\ \bibinfo {volume} {178}\ (\bibinfo
  {publisher} {Elsevier},\ \bibinfo {year} {2013})\ pp.\ \bibinfo {pages}
  {147--247}\BibitemShut {NoStop}%
\bibitem [{\citenamefont {Zhao}\ \emph {et~al.}(2021)\citenamefont {Zhao},
  \citenamefont {Zhang}, \citenamefont {Li}, \citenamefont {Liu}, \citenamefont
  {Miao}, \citenamefont {Shi},\ and\ \citenamefont {Zhao}}]{zhao2021brief}%
  \BibitemOpen
  \bibfield  {author} {\bibinfo {author} {\bibfnamefont {Y.}~\bibnamefont
  {Zhao}}, \bibinfo {author} {\bibfnamefont {J.}~\bibnamefont {Zhang}},
  \bibinfo {author} {\bibfnamefont {J.}~\bibnamefont {Li}}, \bibinfo {author}
  {\bibfnamefont {S.}~\bibnamefont {Liu}}, \bibinfo {author} {\bibfnamefont
  {P.}~\bibnamefont {Miao}}, \bibinfo {author} {\bibfnamefont {Y.}~\bibnamefont
  {Shi}}, \ and\ \bibinfo {author} {\bibfnamefont {E.}~\bibnamefont {Zhao}},\
  }\href@noop {} {\bibfield  {journal} {\bibinfo  {journal} {Measurement
  Science and Technology}\ }\textbf {\bibinfo {volume} {32}},\ \bibinfo {pages}
  {042002} (\bibinfo {year} {2021})}\BibitemShut {NoStop}%
\bibitem [{\citenamefont {Tian}\ \emph {et~al.}(2022)\citenamefont {Tian},
  \citenamefont {Wen}, \citenamefont {Li}, \citenamefont {Ju}, \citenamefont
  {Tang},\ and\ \citenamefont {Xiong}}]{tian2022pfmd}%
  \BibitemOpen
  \bibfield  {author} {\bibinfo {author} {\bibfnamefont {B.}~\bibnamefont
  {Tian}}, \bibinfo {author} {\bibfnamefont {S.}~\bibnamefont {Wen}}, \bibinfo
  {author} {\bibfnamefont {X.}~\bibnamefont {Li}}, \bibinfo {author}
  {\bibfnamefont {J.}~\bibnamefont {Ju}}, \bibinfo {author} {\bibfnamefont
  {J.}~\bibnamefont {Tang}}, \ and\ \bibinfo {author} {\bibfnamefont
  {N.}~\bibnamefont {Xiong}},\ }\href@noop {} {\bibfield  {journal} {\bibinfo
  {journal} {Applied Sciences}\ }\textbf {\bibinfo {volume} {12}},\ \bibinfo
  {pages} {10735} (\bibinfo {year} {2022})}\BibitemShut {NoStop}%
\bibitem [{\citenamefont {Brkovic}\ \emph {et~al.}(1994)\citenamefont
  {Brkovic}, \citenamefont {Pietkiewicz},\ and\ \citenamefont
  {Cuk}}]{brkovic1994novel}%
  \BibitemOpen
  \bibfield  {author} {\bibinfo {author} {\bibfnamefont {M.}~\bibnamefont
  {Brkovic}}, \bibinfo {author} {\bibfnamefont {A.}~\bibnamefont
  {Pietkiewicz}}, \ and\ \bibinfo {author} {\bibfnamefont {S.}~\bibnamefont
  {Cuk}},\ }in\ \href@noop {} {\emph {\bibinfo {booktitle} {Proceedings of
  Intelec 94}}}\ (\bibinfo {organization} {IEEE},\ \bibinfo {year} {1994})\
  pp.\ \bibinfo {pages} {155--162}\BibitemShut {NoStop}%
\bibitem [{\citenamefont {Emig}(2008)}]{Emig_2008}%
  \BibitemOpen
  \bibfield  {author} {\bibinfo {author} {\bibfnamefont {T.}~\bibnamefont
  {Emig}},\ }\href {\doibase 10.1088/1742-5468/2008/04/p04007} {\bibfield
  {journal} {\bibinfo  {journal} {Journal of Statistical Mechanics: Theory and
  Experiment}\ }\textbf {\bibinfo {volume} {2008}},\ \bibinfo {pages} {P04007}
  (\bibinfo {year} {2008})}\BibitemShut {NoStop}%
\bibitem [{\citenamefont {Ess{\'e}n}\ and\ \citenamefont
  {Fiolhais}(2012)}]{essen2012meissner}%
  \BibitemOpen
  \bibfield  {author} {\bibinfo {author} {\bibfnamefont {H.}~\bibnamefont
  {Ess{\'e}n}}\ and\ \bibinfo {author} {\bibfnamefont {M.~C.}\ \bibnamefont
  {Fiolhais}},\ }\href@noop {} {\bibfield  {journal} {\bibinfo  {journal}
  {American Journal of Physics}\ }\textbf {\bibinfo {volume} {80}},\ \bibinfo
  {pages} {164} (\bibinfo {year} {2012})}\BibitemShut {NoStop}%
\bibitem [{\citenamefont {Iverson}\ and\ \citenamefont
  {Kuhl}(1995)}]{iverson1995mapping}%
  \BibitemOpen
  \bibfield  {author} {\bibinfo {author} {\bibfnamefont {P.}~\bibnamefont
  {Iverson}}\ and\ \bibinfo {author} {\bibfnamefont {P.~K.}\ \bibnamefont
  {Kuhl}},\ }\href@noop {} {\bibfield  {journal} {\bibinfo  {journal} {The
  Journal of the Acoustical Society of America}\ }\textbf {\bibinfo {volume}
  {97}},\ \bibinfo {pages} {553} (\bibinfo {year} {1995})}\BibitemShut
  {NoStop}%
\bibitem [{\citenamefont {Naitou}\ \emph {et~al.}(1979)\citenamefont {Naitou},
  \citenamefont {Tokuda},\ and\ \citenamefont {Kamimura}}]{naitou1979boundary}%
  \BibitemOpen
  \bibfield  {author} {\bibinfo {author} {\bibfnamefont {H.}~\bibnamefont
  {Naitou}}, \bibinfo {author} {\bibfnamefont {S.}~\bibnamefont {Tokuda}}, \
  and\ \bibinfo {author} {\bibfnamefont {T.}~\bibnamefont {Kamimura}},\
  }\href@noop {} {\bibfield  {journal} {\bibinfo  {journal} {Journal of
  Computational Physics}\ }\textbf {\bibinfo {volume} {33}},\ \bibinfo {pages}
  {86} (\bibinfo {year} {1979})}\BibitemShut {NoStop}%
\bibitem [{\citenamefont {Charitat}\ and\ \citenamefont
  {Graner}(2003)}]{Charitat_2003}%
  \BibitemOpen
  \bibfield  {author} {\bibinfo {author} {\bibfnamefont {T.}~\bibnamefont
  {Charitat}}\ and\ \bibinfo {author} {\bibfnamefont {F.}~\bibnamefont
  {Graner}},\ }\href {\doibase 10.1088/0143-0807/24/3/306} {\bibfield
  {journal} {\bibinfo  {journal} {European Journal of Physics}\ }\textbf
  {\bibinfo {volume} {24}},\ \bibinfo {pages} {267} (\bibinfo {year}
  {2003})}\BibitemShut {NoStop}%
\bibitem [{\citenamefont {Oliveira}\ and\ \citenamefont
  {Miranda}(2001)}]{oliveira2001biot}%
  \BibitemOpen
  \bibfield  {author} {\bibinfo {author} {\bibfnamefont {M.~H.}\ \bibnamefont
  {Oliveira}}\ and\ \bibinfo {author} {\bibfnamefont {J.~A.}\ \bibnamefont
  {Miranda}},\ }\href@noop {} {\bibfield  {journal} {\bibinfo  {journal}
  {European Journal of Physics}\ }\textbf {\bibinfo {volume} {22}},\ \bibinfo
  {pages} {31} (\bibinfo {year} {2001})}\BibitemShut {NoStop}%
\bibitem [{\citenamefont {Albani}\ and\ \citenamefont
  {Bernardi}(1974)}]{1128246}%
  \BibitemOpen
  \bibfield  {author} {\bibinfo {author} {\bibfnamefont {M.}~\bibnamefont
  {Albani}}\ and\ \bibinfo {author} {\bibfnamefont {P.}~\bibnamefont
  {Bernardi}},\ }\href {\doibase 10.1109/TMTT.1974.1128246} {\bibfield
  {journal} {\bibinfo  {journal} {IEEE Transactions on Microwave Theory and
  Techniques}\ }\textbf {\bibinfo {volume} {22}},\ \bibinfo {pages} {446}
  (\bibinfo {year} {1974})}\BibitemShut {NoStop}%
\bibitem [{\citenamefont {{\'A}rnason}(1999)}]{arnason1999consistent}%
  \BibitemOpen
  \bibfield  {author} {\bibinfo {author} {\bibfnamefont {K.}~\bibnamefont
  {{\'A}rnason}},\ }in\ \href@noop {} {\emph {\bibinfo {booktitle}
  {Three-Dimensional Electromagnetics}}}\ (\bibinfo  {publisher} {Society of
  Exploration Geophysicists},\ \bibinfo {year} {1999})\ pp.\ \bibinfo {pages}
  {103--118}\BibitemShut {NoStop}%
\bibitem [{\citenamefont {Bossavit}(2005)}]{BOSSAVIT2005105}%
  \BibitemOpen
  \bibfield  {author} {\bibinfo {author} {\bibfnamefont {A.}~\bibnamefont
  {Bossavit}},\ }in\ \href {\doibase
  https://doi.org/10.1016/S1570-8659(04)13002-0} {\emph {\bibinfo {booktitle}
  {Numerical Methods in Electromagnetics}}},\ \bibinfo {series} {Handbook of
  Numerical Analysis}, Vol.~\bibinfo {volume} {13}\ (\bibinfo  {publisher}
  {Elsevier},\ \bibinfo {year} {2005})\ pp.\ \bibinfo {pages}
  {105--197}\BibitemShut {NoStop}%
\bibitem [{\citenamefont {Bardeen}(1955)}]{bardeen1955theory}%
  \BibitemOpen
  \bibfield  {author} {\bibinfo {author} {\bibfnamefont {J.}~\bibnamefont
  {Bardeen}},\ }\href@noop {} {\bibfield  {journal} {\bibinfo  {journal}
  {Physical Review}\ }\textbf {\bibinfo {volume} {97}},\ \bibinfo {pages}
  {1724} (\bibinfo {year} {1955})}\BibitemShut {NoStop}%
\bibitem [{\citenamefont {Schafroth}(1958)}]{schafroth1958remarks}%
  \BibitemOpen
  \bibfield  {author} {\bibinfo {author} {\bibfnamefont {M.}~\bibnamefont
  {Schafroth}},\ }\href@noop {} {\bibfield  {journal} {\bibinfo  {journal}
  {Physical Review}\ }\textbf {\bibinfo {volume} {111}},\ \bibinfo {pages} {72}
  (\bibinfo {year} {1958})}\BibitemShut {NoStop}%
\bibitem [{\citenamefont {Giuliani}(2008)}]{Giuliani_2008}%
  \BibitemOpen
  \bibfield  {author} {\bibinfo {author} {\bibfnamefont {G.}~\bibnamefont
  {Giuliani}},\ }\href {\doibase 10.1209/0295-5075/81/60002} {\bibfield
  {journal} {\bibinfo  {journal} {{EPL} (Europhysics Letters)}\ }\textbf
  {\bibinfo {volume} {81}},\ \bibinfo {pages} {60002} (\bibinfo {year}
  {2008})}\BibitemShut {NoStop}%
\bibitem [{\citenamefont {Galili}\ \emph {et~al.}(2006)\citenamefont {Galili},
  \citenamefont {Kaplan},\ and\ \citenamefont {Lehavi}}]{galili2006teaching}%
  \BibitemOpen
  \bibfield  {author} {\bibinfo {author} {\bibfnamefont {I.}~\bibnamefont
  {Galili}}, \bibinfo {author} {\bibfnamefont {D.}~\bibnamefont {Kaplan}}, \
  and\ \bibinfo {author} {\bibfnamefont {Y.}~\bibnamefont {Lehavi}},\
  }\href@noop {} {\bibfield  {journal} {\bibinfo  {journal} {American journal
  of physics}\ }\textbf {\bibinfo {volume} {74}},\ \bibinfo {pages} {337}
  (\bibinfo {year} {2006})}\BibitemShut {NoStop}%
\bibitem [{\citenamefont {Uranga}\ \emph {et~al.}(2004)\citenamefont {Uranga},
  \citenamefont {Lago}, \citenamefont {Navarro},\ and\ \citenamefont
  {Barniol}}]{1328930}%
  \BibitemOpen
  \bibfield  {author} {\bibinfo {author} {\bibfnamefont {A.}~\bibnamefont
  {Uranga}}, \bibinfo {author} {\bibfnamefont {N.}~\bibnamefont {Lago}},
  \bibinfo {author} {\bibfnamefont {X.}~\bibnamefont {Navarro}}, \ and\
  \bibinfo {author} {\bibfnamefont {N.}~\bibnamefont {Barniol}},\ }in\ \href
  {\doibase 10.1109/ISCAS.2004.1328930} {\emph {\bibinfo {booktitle} {2004 IEEE
  International Symposium on Circuits and Systems (IEEE Cat. No.04CH37512)}}},\
  Vol.~\bibinfo {volume} {4}\ (\bibinfo {year} {2004})\ pp.\ \bibinfo {pages}
  {IV--21}\BibitemShut {NoStop}%
\bibitem [{\citenamefont {GUIDE}()}]{guide18active}%
  \BibitemOpen
  \bibfield  {author} {\bibinfo {author} {\bibfnamefont {S.}~\bibnamefont
  {GUIDE}},\ }\href@noop {} {\bibfield  {journal} {\bibinfo  {journal} {Power}\
  }\textbf {\bibinfo {volume} {18}},\ \bibinfo {pages} {11}}\BibitemShut
  {NoStop}%
\bibitem [{\citenamefont {Schilling}\ and\ \citenamefont
  {Shih}(2019)}]{schilling2019parametric}%
  \BibitemOpen
  \bibfield  {author} {\bibinfo {author} {\bibfnamefont {P.}~\bibnamefont
  {Schilling}}\ and\ \bibinfo {author} {\bibfnamefont {R.}~\bibnamefont
  {Shih}},\ }\href@noop {} {\emph {\bibinfo {title} {Parametric Modeling with
  SOLIDWORKS 2019}}}\ (\bibinfo  {publisher} {SDC Publications},\ \bibinfo
  {year} {2019})\BibitemShut {NoStop}%
\bibitem [{\citenamefont {Akin}(2010)}]{akin2010finite}%
  \BibitemOpen
  \bibfield  {author} {\bibinfo {author} {\bibfnamefont {J.~E.}\ \bibnamefont
  {Akin}},\ }\href@noop {} {\emph {\bibinfo {title} {Finite element analysis
  concepts: via SolidWorks}}}\ (\bibinfo  {publisher} {World Scientific},\
  \bibinfo {year} {2010})\BibitemShut {NoStop}%
\bibitem [{\citenamefont {Avizienis}\ and\ \citenamefont
  {Kelly}(1984)}]{avizienis1984fault}%
  \BibitemOpen
  \bibfield  {author} {\bibinfo {author} {\bibfnamefont {A.}~\bibnamefont
  {Avizienis}}\ and\ \bibinfo {author} {\bibfnamefont {J.~P.}\ \bibnamefont
  {Kelly}},\ }\href@noop {} {\bibfield  {journal} {\bibinfo  {journal}
  {Computer}\ }\textbf {\bibinfo {volume} {17}},\ \bibinfo {pages} {67}
  (\bibinfo {year} {1984})}\BibitemShut {NoStop}%
\bibitem [{\citenamefont {Bertoni}\ \emph {et~al.}(2003)\citenamefont
  {Bertoni}, \citenamefont {Breveglieri}, \citenamefont {Koren}, \citenamefont
  {Maistri},\ and\ \citenamefont {Piuri}}]{bertoni2003error}%
  \BibitemOpen
  \bibfield  {author} {\bibinfo {author} {\bibfnamefont {G.}~\bibnamefont
  {Bertoni}}, \bibinfo {author} {\bibfnamefont {L.}~\bibnamefont
  {Breveglieri}}, \bibinfo {author} {\bibfnamefont {I.}~\bibnamefont {Koren}},
  \bibinfo {author} {\bibfnamefont {P.}~\bibnamefont {Maistri}}, \ and\
  \bibinfo {author} {\bibfnamefont {V.}~\bibnamefont {Piuri}},\ }\href@noop {}
  {\bibfield  {journal} {\bibinfo  {journal} {IEEE transactions on Computers}\
  }\textbf {\bibinfo {volume} {52}},\ \bibinfo {pages} {492} (\bibinfo {year}
  {2003})}\BibitemShut {NoStop}%
\bibitem [{\citenamefont {Pan}\ \emph {et~al.}(2020)\citenamefont {Pan},
  \citenamefont {Lin}, \citenamefont {Li}, \citenamefont {Li}, \citenamefont
  {Jin}, \citenamefont {Sun},\ and\ \citenamefont {Liu}}]{8653466}%
  \BibitemOpen
  \bibfield  {author} {\bibinfo {author} {\bibfnamefont {D.}~\bibnamefont
  {Pan}}, \bibinfo {author} {\bibfnamefont {S.}~\bibnamefont {Lin}}, \bibinfo
  {author} {\bibfnamefont {L.}~\bibnamefont {Li}}, \bibinfo {author}
  {\bibfnamefont {J.}~\bibnamefont {Li}}, \bibinfo {author} {\bibfnamefont
  {Y.}~\bibnamefont {Jin}}, \bibinfo {author} {\bibfnamefont {Z.}~\bibnamefont
  {Sun}}, \ and\ \bibinfo {author} {\bibfnamefont {T.}~\bibnamefont {Liu}},\
  }\href {\doibase 10.1109/TIE.2019.2899544} {\bibfield  {journal} {\bibinfo
  {journal} {IEEE Transactions on Industrial Electronics}\ }\textbf {\bibinfo
  {volume} {67}},\ \bibinfo {pages} {1348} (\bibinfo {year}
  {2020})}\BibitemShut {NoStop}%
\bibitem [{\citenamefont {Shou}\ \emph {et~al.}(2010)\citenamefont {Shou},
  \citenamefont {Xia}, \citenamefont {Liu}, \citenamefont {Zhu}, \citenamefont
  {Li},\ and\ \citenamefont {Crozier}}]{5352342}%
  \BibitemOpen
  \bibfield  {author} {\bibinfo {author} {\bibfnamefont {G.}~\bibnamefont
  {Shou}}, \bibinfo {author} {\bibfnamefont {L.}~\bibnamefont {Xia}}, \bibinfo
  {author} {\bibfnamefont {F.}~\bibnamefont {Liu}}, \bibinfo {author}
  {\bibfnamefont {M.}~\bibnamefont {Zhu}}, \bibinfo {author} {\bibfnamefont
  {Y.}~\bibnamefont {Li}}, \ and\ \bibinfo {author} {\bibfnamefont
  {S.}~\bibnamefont {Crozier}},\ }\href {\doibase 10.1109/TMAG.2009.2037753}
  {\bibfield  {journal} {\bibinfo  {journal} {IEEE Transactions on Magnetics}\
  }\textbf {\bibinfo {volume} {46}},\ \bibinfo {pages} {1052} (\bibinfo {year}
  {2010})}\BibitemShut {NoStop}%
\bibitem [{\citenamefont {Evans}\ \emph {et~al.}(2005)\citenamefont {Evans},
  \citenamefont {Moyer}, \citenamefont {Watkins}, \citenamefont {Thomas},
  \citenamefont {Osborne}, \citenamefont {Boedo}, \citenamefont
  {Fenstermacher}, \citenamefont {Finken}, \citenamefont {Groebner},
  \citenamefont {Groth}, \citenamefont {Harris}, \citenamefont {Jackson},
  \citenamefont {Haye}, \citenamefont {Lasnier}, \citenamefont {Schaffer},
  \citenamefont {Wang},\ and\ \citenamefont {Zeng}}]{EVANS2005691}%
  \BibitemOpen
  \bibfield  {author} {\bibinfo {author} {\bibfnamefont {T.}~\bibnamefont
  {Evans}}, \bibinfo {author} {\bibfnamefont {R.}~\bibnamefont {Moyer}},
  \bibinfo {author} {\bibfnamefont {J.}~\bibnamefont {Watkins}}, \bibinfo
  {author} {\bibfnamefont {P.}~\bibnamefont {Thomas}}, \bibinfo {author}
  {\bibfnamefont {T.}~\bibnamefont {Osborne}}, \bibinfo {author} {\bibfnamefont
  {J.}~\bibnamefont {Boedo}}, \bibinfo {author} {\bibfnamefont
  {M.}~\bibnamefont {Fenstermacher}}, \bibinfo {author} {\bibfnamefont
  {K.}~\bibnamefont {Finken}}, \bibinfo {author} {\bibfnamefont
  {R.}~\bibnamefont {Groebner}}, \bibinfo {author} {\bibfnamefont
  {M.}~\bibnamefont {Groth}}, \bibinfo {author} {\bibfnamefont
  {J.}~\bibnamefont {Harris}}, \bibinfo {author} {\bibfnamefont
  {G.}~\bibnamefont {Jackson}}, \bibinfo {author} {\bibfnamefont {R.~L.}\
  \bibnamefont {Haye}}, \bibinfo {author} {\bibfnamefont {C.}~\bibnamefont
  {Lasnier}}, \bibinfo {author} {\bibfnamefont {M.}~\bibnamefont {Schaffer}},
  \bibinfo {author} {\bibfnamefont {G.}~\bibnamefont {Wang}}, \ and\ \bibinfo
  {author} {\bibfnamefont {L.}~\bibnamefont {Zeng}},\ }\href {\doibase
  https://doi.org/10.1016/j.jnucmat.2004.10.062} {\bibfield  {journal}
  {\bibinfo  {journal} {Journal of Nuclear Materials}\ }\textbf {\bibinfo
  {volume} {337-339}},\ \bibinfo {pages} {691} (\bibinfo {year} {2005})},\
  \bibinfo {note} {pSI-16}\BibitemShut {NoStop}%
\bibitem [{\citenamefont {Yu}\ and\ \citenamefont
  {Girshick}(1991)}]{yu1991modeling}%
  \BibitemOpen
  \bibfield  {author} {\bibinfo {author} {\bibfnamefont {B.~W.}\ \bibnamefont
  {Yu}}\ and\ \bibinfo {author} {\bibfnamefont {S.~L.}\ \bibnamefont
  {Girshick}},\ }\href@noop {} {\bibfield  {journal} {\bibinfo  {journal}
  {Journal of applied physics}\ }\textbf {\bibinfo {volume} {69}},\ \bibinfo
  {pages} {656} (\bibinfo {year} {1991})}\BibitemShut {NoStop}%
\bibitem [{\citenamefont {Anderson}\ and\ \citenamefont
  {Matthias}(1964)}]{anderson1964superconductivity}%
  \BibitemOpen
  \bibfield  {author} {\bibinfo {author} {\bibfnamefont {P.}~\bibnamefont
  {Anderson}}\ and\ \bibinfo {author} {\bibfnamefont {B.}~\bibnamefont
  {Matthias}},\ }\href@noop {} {\bibfield  {journal} {\bibinfo  {journal}
  {Science}\ }\textbf {\bibinfo {volume} {144}},\ \bibinfo {pages} {373}
  (\bibinfo {year} {1964})}\BibitemShut {NoStop}%
\bibitem [{\citenamefont {Kaloper}\ and\ \citenamefont
  {Lawrence}(2017)}]{kaloper2017london}%
  \BibitemOpen
  \bibfield  {author} {\bibinfo {author} {\bibfnamefont {N.}~\bibnamefont
  {Kaloper}}\ and\ \bibinfo {author} {\bibfnamefont {A.}~\bibnamefont
  {Lawrence}},\ }\href@noop {} {\bibfield  {journal} {\bibinfo  {journal}
  {Physical Review D}\ }\textbf {\bibinfo {volume} {95}},\ \bibinfo {pages}
  {063526} (\bibinfo {year} {2017})}\BibitemShut {NoStop}%
\bibitem [{\citenamefont {Lax}\ and\ \citenamefont
  {Nelson}(1976)}]{lax1976maxwell}%
  \BibitemOpen
  \bibfield  {author} {\bibinfo {author} {\bibfnamefont {M.}~\bibnamefont
  {Lax}}\ and\ \bibinfo {author} {\bibfnamefont {D.}~\bibnamefont {Nelson}},\
  }\href@noop {} {\bibfield  {journal} {\bibinfo  {journal} {Physical Review
  B}\ }\textbf {\bibinfo {volume} {13}},\ \bibinfo {pages} {1777} (\bibinfo
  {year} {1976})}\BibitemShut {NoStop}%
\bibitem [{\citenamefont {Assous}\ \emph {et~al.}(1993)\citenamefont {Assous},
  \citenamefont {Degond}, \citenamefont {Heintze}, \citenamefont {Raviart},\
  and\ \citenamefont {Segre}}]{ASSOUS1993222}%
  \BibitemOpen
  \bibfield  {author} {\bibinfo {author} {\bibfnamefont {F.}~\bibnamefont
  {Assous}}, \bibinfo {author} {\bibfnamefont {P.}~\bibnamefont {Degond}},
  \bibinfo {author} {\bibfnamefont {E.}~\bibnamefont {Heintze}}, \bibinfo
  {author} {\bibfnamefont {P.}~\bibnamefont {Raviart}}, \ and\ \bibinfo
  {author} {\bibfnamefont {J.}~\bibnamefont {Segre}},\ }\href {\doibase
  https://doi.org/10.1006/jcph.1993.1214} {\bibfield  {journal} {\bibinfo
  {journal} {Journal of Computational Physics}\ }\textbf {\bibinfo {volume}
  {109}},\ \bibinfo {pages} {222} (\bibinfo {year} {1993})}\BibitemShut
  {NoStop}%
\bibitem [{\citenamefont {Gurevich}\ and\ \citenamefont
  {Laiho}(1993)}]{gurevich1993photomagnetism}%
  \BibitemOpen
  \bibfield  {author} {\bibinfo {author} {\bibfnamefont {V.}~\bibnamefont
  {Gurevich}}\ and\ \bibinfo {author} {\bibfnamefont {R.}~\bibnamefont
  {Laiho}},\ }\href@noop {} {\bibfield  {journal} {\bibinfo  {journal}
  {Physical Review B}\ }\textbf {\bibinfo {volume} {48}},\ \bibinfo {pages}
  {8307} (\bibinfo {year} {1993})}\BibitemShut {NoStop}%
\bibitem [{\citenamefont {Gurevich}\ \emph {et~al.}(1992)\citenamefont
  {Gurevich}, \citenamefont {Laiho},\ and\ \citenamefont
  {Lashkul}}]{gurevich1992photomagnetism}%
  \BibitemOpen
  \bibfield  {author} {\bibinfo {author} {\bibfnamefont {V.}~\bibnamefont
  {Gurevich}}, \bibinfo {author} {\bibfnamefont {R.}~\bibnamefont {Laiho}}, \
  and\ \bibinfo {author} {\bibfnamefont {A.}~\bibnamefont {Lashkul}},\
  }\href@noop {} {\bibfield  {journal} {\bibinfo  {journal} {Physical review
  letters}\ }\textbf {\bibinfo {volume} {69}},\ \bibinfo {pages} {180}
  (\bibinfo {year} {1992})}\BibitemShut {NoStop}%
\end{thebibliography}%
\end{document}